\documentclass[conference]{IEEEtran}
\IEEEoverridecommandlockouts
\usepackage{cite}
\usepackage{amsmath,amssymb,amsfonts}
\usepackage{algorithmic}
\usepackage{graphicx}
\usepackage{textcomp}
\usepackage{xcolor}
\usepackage{multirow}
\usepackage{subfig} 
\usepackage{booktabs}
\def\BibTeX{{\rm B\kern-.05em{\sc i\kern-.025em b}\kern-.08em
    T\kern-.1667em\lower.7ex\hbox{E}\kern-.125emX}}
\begin{document}

%\title{Time Series Face Portrait for Ball-Bearing Failure Detection using DCGAN Networks\\
%}
\title{Digital Twin Enabled Smart Control Engineering as an Industrial AI: A New Framework and A Case Study}

\author{\IEEEauthorblockN{Jairo Viola and YangQuan Chen}
\IEEEauthorblockA{\textit{MESA Lab}\\
\textit{University of California, Merced}\\
Merced, CA, USA \\
{(jviola,ychen53)}@ucmerced.edu}
%\and
%\IEEEauthorblockN{Jing Wang}
%\IEEEauthorblockA{\textit{College of Automation} \\
%\textit{Beijing University of Chemical Technology}\\
%Beijing, China \\
%jwang@mail.buct.edu.cn}
}

\maketitle

\begin{abstract}
	
In the way towards Industry 4.0, the complexity of the industrial systems increases due to the presence of multiple agents, Cyber-Physical Systems, distributed sensing, and big data introducing unknown dynamics that affect the production goals of the manufacturing processes. Thus, Digital Twin is a breaking technology corresponding to the capacity of developing a virtual representation of any complex system in order to perform design, analysis, and behavior prediction tasks that enhance the understanding of these systems through new enabling capabilities like real-time analytics, parallel sensing, or Smart Control Engineering. In this paper, a novel framework is proposed for the design and implementation of Digital Twin applications to the development of Smart Control Engineering. The steps of this framework involve system documentation, multidomain simulation, behavioral matching, and real-time monitoring. This framework is applied to develop the Digital Twin for a real-time vision feedback infrared temperature uniformity control. The obtained results show that Digital Twin is a fundamental part of the transformation into Industry 4.0.\\
\end{abstract}


\section{Introduction}
Modern manufacturing processes require not only high-quality standards but also enhanced robustness and autonomy in order to achieve production objectives. So that, in the Industry 4.0, the convergence of multiple breaking technologies like Deep Learning, Artificial Intelligence, Big Data, real-time analytics is required to introduce newly automated and enabled capabilities to introduce smartness into manufacturing processes. The application of these technologies into the industrial environment is known as Industrial Artificial Intelligence (IAI), which is focused on helping an enterprise to monitor, optimize or control the behavior of these systems, improving its efficiency and performance \cite{IAIDefinition}. Thus, multiple approaches of IAI can be found in the literature to deal with specific problems in the industry like predictive maintenance, supply chain, quality control, fault detection, and isolation, among others \cite{Zhang2019,Li2020,Hao2019,Mao2019,Dong2019,Li2019}.
\par
From a classic perspective, the automatic control, which is present in almost all the industrial processes, requires an output feedback signal from the system to be controlled, that is compared with a desired setpoint, producing an error signal processed by a control algorithm that generates a control action to stabilize the system and accomplish some closed-loop desired specifications. In the IAI and industry 4.0 framework, the integration of breaking technologies offers to the control system designers new tools to improve the control performance like information-rich systems and machine learning techniques, transforming the classic Control Engineering into Smart Control Engineering \cite{CPCChen}.
\par
Thus, a unified, detailed, and realistic representation of the systems on Industry 4.0 is required in order to leverage the convergence of breaking technologies with IAI capabilities to enable Smart Control Engineering. So that, Digital Twin (DT) offers the possibility of having a highly detailed and realistic virtual model of any system that replicates its real behavior based on the system physics and data-driven models of each subsystem. In addition, the subsystems interaction and other system features can be considered, which are harder to incorporate from a classic Control Engineering. In the recent years different Digital Twin applications has been performed for systems like drones, UAV \cite{Guivarch2019}, power systems \cite{XieSG}, CNC machines \cite{Luo2018} or complex models like smart transportation \cite{Zhu2016}, smart grid \cite{Claessen2012,Angel2019} , or smart cities \cite{Farsi2020}. However, a new framework is required to develop Digital Twin applications for Smart Control Engineering.
\par
This paper presents a novel framework for developing Digital Twin applications for industrial process control applications towards a Smart Control Engineering. This framework consists of five steps: Target system definition, System Documentation, Multidomain simulation, Digital Twin assembly, and Behavioral matching and DT validation and deployment. This framework is applied for designing the Digital Twin of a real-time vision feedback infrared temperature uniformity control. A detailed list of attributes of the system is made to select the critical components of the system that will be incorporated in the DT. The multidomain simulation considers the Electrical, Thermal, and Digital simulation domains that compose the thermal process, using Simscape multiphysics to represent each domain according to its physical constitutive laws. The Behavioral matching is performed using Genetic Algorithms (GA) to determine the unknown parameters at every single element in the multidomain simulation DT. Finally, the DT deployment is performed via a supervisory interface to assess the real-time thermal system performance with the Digital Twin running in parallel. 
\par
The main contribution of this paper is to introduce a new framework for developing Digital Twin applications in process control towards the development of Smart Control Engineering and its application into the Industrial Artificial Intelligence.
\par
The manuscript is structured as follows. Section II presents an overview of Digital Twin and Smart Control Engineering. Sections III and IV show the new framework for Digital Twin applications development as well as the Study case corresponding to the real-time vision feedback infrared temperature uniformity control system. Finally, conclusions and future works presented.

\section{Digital Twin and Smart Control Engineering}
\subsection{What is Digital Twin?}
\begin{figure}[h]
	\centering
	\includegraphics[width=0.45\textwidth, height=0.14\textheight]{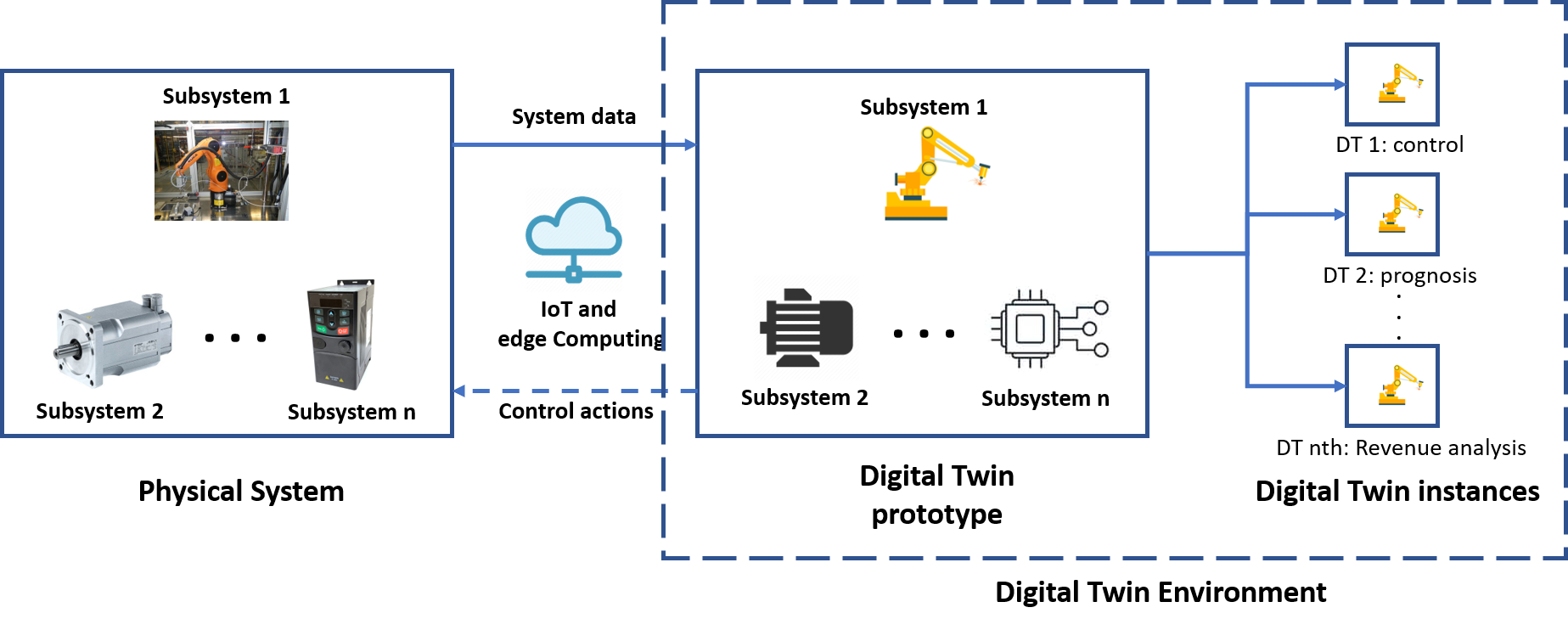}
	\caption{Digital Twin structure}
	\label{DT Concept}
\end{figure}

Digital Twin (DT) can be defined as a precise, virtual copies of machines or systems driven by data collected from sensors in real-time. These sophisticated computer models mirror almost every facet of a product, process, or service \cite{Tao2019Nature}. According to \cite{GartnetInc.2018}, DT is expected to be a \$35.8 Billion USD market for 2025. A graphical description of Digital Twin is presented in Fig.\ref{DT Concept}. As can be observed, the DT model starts from a physical system, which is composed of multiple subsystems like sensors or actuators. Every single element has a virtual representation or prototype inside the Digital Twin environment, corresponding to a multi-domain simulation model built using different physics-based simulation tools integrated to work together to replicate the real system behavior. Likewise, the prototype can have multiple instances for a specific task like control system design, prognosis, Fault simulation, or machine learning training task. Also, the instances can be aggregated if the specific tasks require multiple results based on the same virtual representation. Notice that the DT environment is feed with real-time data from the physical system through the Internet of Things and Edge Computing devices.
\subsection{Digital Twin features and components}
\begin{figure}[h]
	\centering
	\includegraphics[width=0.4\textwidth, height=0.14\textheight]{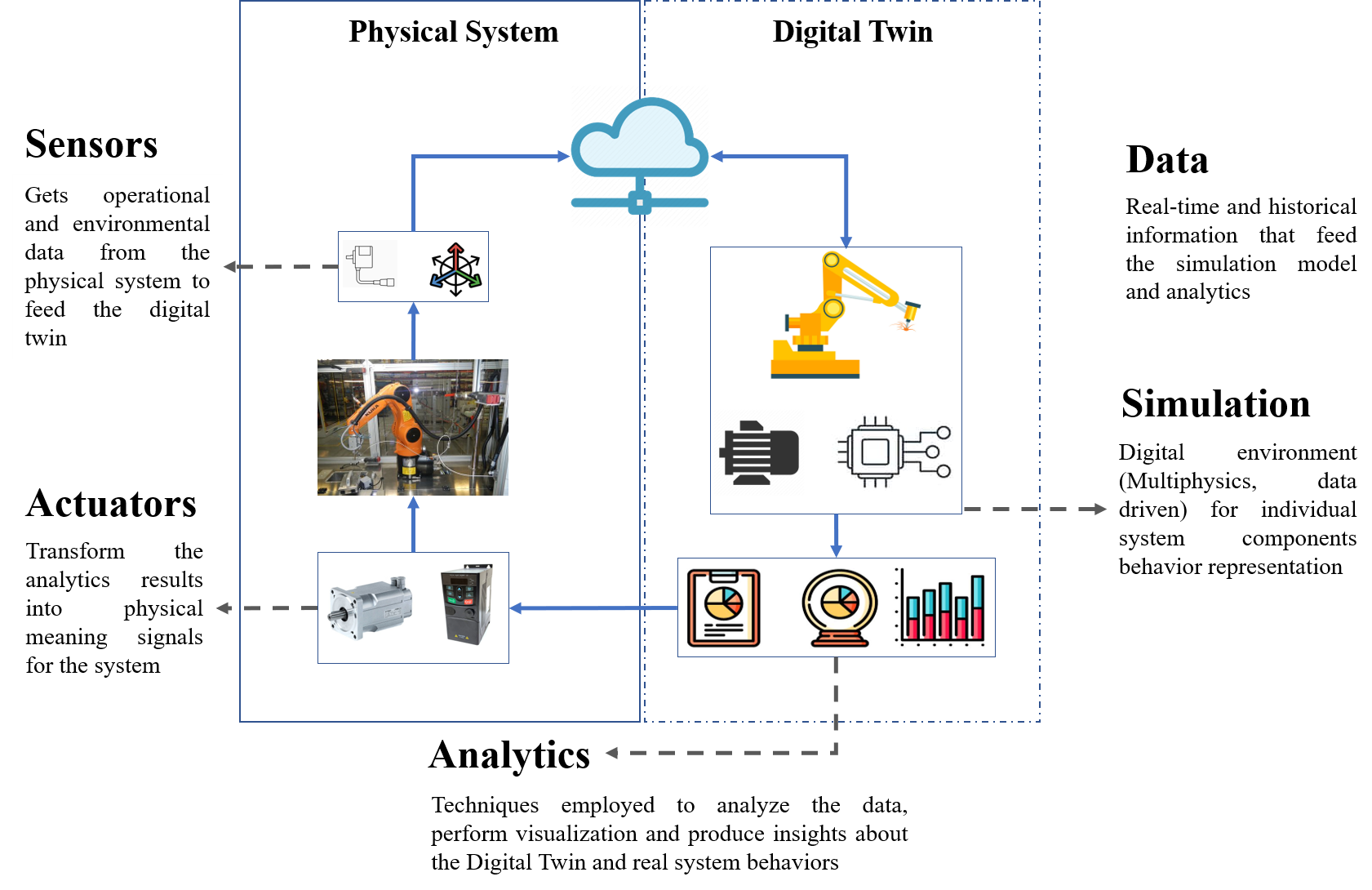}
	\caption{Digital Twin components}
	\label{DTcomponents}
\end{figure}

The main features of a Digital Twin application can be defined as follows:
\begin{itemize}
	\item \textbf{Realizability:} The virtual representation of the physical system is performed through virtual entities or avatars based on a mixed model and data-driven based design.
	\item \textbf{Real-Time system data updating:}The enabling sensing and analytics technologies like IoT and Edge computing allows the DT to update the system data with real-time or historical batch datasets
	\item \textbf{Behavioral Matching:}The system behavior matching relies not only on experience-based knowledge but also in adaptive optimizing techniques feed with the real-time data updating information.
	\item \textbf{Modular Structure:} A DT Application is built by individual subsystem modules developed in multiple multi-physics software and data-driven tools, which are integrated to emulate the physical system behavior.
	\item \textbf{Trackability:} The time evolution of the system can be recorded using different DT model instances as time capsules for the current and previous states of each subsystem as well as the whole DT application.
	\item \textbf{Reprogramability:} DT modules can be reprogramed according to the new system requirements, physical changes, and current system status.
\end{itemize}
On the other hand, the principal components of a Digital Twin are shown in Fig.\ref{DTcomponents}. It can be observed that a DT application is composed of sensors and actuators in the physical system, which collect the system information and apply the control actions over the system. Also, the DT uses the data collected from the system to feed the multidomain simulations and perform analytics not only over the DT simulations but also over the real system to produce recommendations and possible control actions over the system to improve its operation.

\subsection{Smart Control Engineering}
The concept of Smart Control Engineering begins with the conception of Cognitive Process Control (CPC) proposed by Chen \cite{CPCChen}. In this sense, process control can be considered cognitive if the controller architecture is aware of the process of vital signs for healthy runs, performs decision making, and health issues alerting using multiple information sources. Also, the control system is capable of learning from past actions and induced errors (resilience), discovering hidden patterns and anomalous behaviors at multiple time scales. 
\par
These capabilities are supported by information-driven control like Run to Run, Iterative Learning, or Self Optimizing control [cites], sensor-rich information, deep learning techniques, Real-Time Analytics, Edge computing among other breaking technologies. So, the process of Cognition, then decide, then control can be performed integrating a wide picture of the system, creating a smart control system, which according to the Smart and Autonomous Systems (S\&AS) program  of NSF (Natural Sciences Foundation) \cite{NSF2018} has the following characteristics:
\begin{itemize}
	\item \textbf{Cognizant:} the system is aware of capabilities and limitations to face the dynamic changes and variability.
	\item \textbf{Taskable:} The system is capable of handle high-level, often vague instructions based on the automated stimulus or human commands.
	\item \textbf{Reflective:} A smart system is able to learn from previous experience to improve its performance.
	\item \textbf{Knowledge Rich:} All the reasoning processes make by a smart control system are performed over a diverse body of knowledge from a rich environment and sensor-based information.
	\item \textbf{Ethical:} The smart behavior of a system adheres to a system of societal and legal norms.
\end{itemize}
Thus, a new control framework which combines cognition capabilities with automatic control to introduce intelligence for Industry 4.0 manufacturing processes defined as Smart Control Engineering, which is supported by Industrial artificial intelligence, breaking technologies like Cloud computing, Deep learning, Data analytics, Big data among others. However, all these new capabilities and control paradigm requires realistic virtual representations of the physical system where is possible for the control engineers incorporate all the smartness features as well as perform complex real-time analysis over the system.
\par
In this scenario, Digital Twin became a relevant tool to leverage all the features of CPC and smart control engineering via real-time updated virtual representations of the system, considering that is difficult to represent fault behaviors and undesired situations in real life due to the risk associated and high cost of running these tests. Based on the breaking technologies involved in Digital Twin, a set of enabled capabilities can be leveraged towards Smart Control Engineering applications. Among these new capabilities can be found the Smart control system design \cite{ChenMAD}, Control performance assessment \cite{domanski2019control}, Self optimizing control \cite{SKOGESTAD2000487}, Prognosis and fault detection \cite{Isermann2006}, parallel sensing and control \cite{Kang2017} among others.

\section{Development framework for Digital Twin applications}
\begin{figure}[h]
	\centering
	\includegraphics[width=0.35\textwidth, height=0.08\textheight]{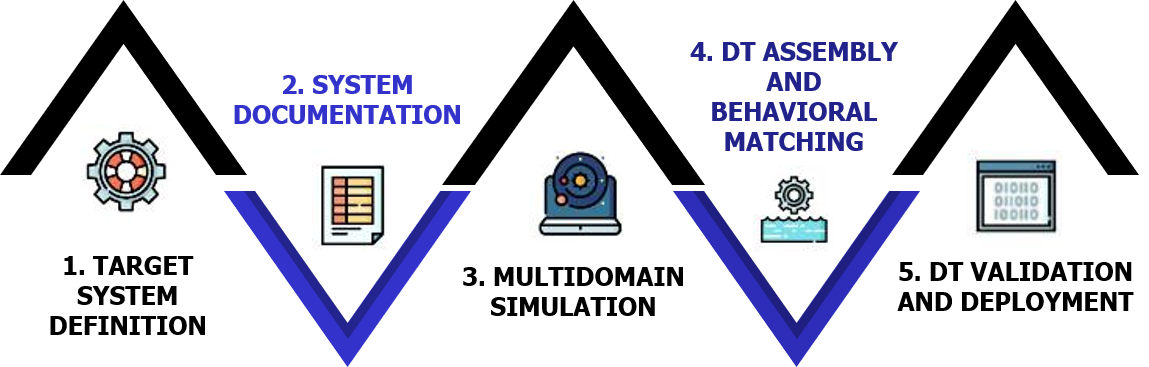}
	\caption{Digital Twin framework}
	\label{DT Framework}
\end{figure}

Figure \ref{DT Framework} presents a methodological framework for the successful implementation of a DT. The framework is composed of five steps corresponding to the target system definition, system documentation, Multidomain simulation, DT assembly and behavioral matching, and the DT evaluation and deployment. The main advantage of this methodology is that one is applied; the Digital Twin is ready for the implementation of the multiple enabled capabilities listed above.

\subsection{Step 1: Target system definition}
This step is focused on recognizing the current status of the physical system to be replicated via Digital Twin with two possible scenarios. The first one is if the physical system is in a conceptual design stage. In this scenario, building a Digital Twin of the system is a preliminary step for physical sizing as well as emulate its real behavior as closer to reality as possible. In this case, using CAD/CAM tools and design is the best choice to create a real system representation.
Notice that after the real system is assembled and operative, the Digital Twin is updatable with the real system information to be aware of the real behavior of the system.
\par
On the other hand, the second scenario is when the physical system is already operative, performing the desired tasks. In this case, even if a previous CAD/CAM model is not available, the Digital Twin can be also be built using as reference the current system configuration using the system big data, information, and experimental knowledge provided by designers, engineers, and operators. In both scenarios, the proposed framework can be employed to create the Digital Twin of a system. In this paper, the second scenario is applied to the study case.

\subsection{Step 2: System Documentation}
Once the target and operating scenario of the system to be replicated by the Digital Twin is defined, the following step consists of collect all the available information of the system to create the most accurate representation. This relevant information includes the control algorithms employed (PID, MPC, State Space) and its digital implementation, Sensors and actuators datasheets, Troubleshooting and problem records, Cumulative experience of the system engineers and operators, and the system data streams.  Notice that most of each subsystem data is available and can be acquired via sensors, virtual metrology, indirect measurements, or state estimators. In addition, the information about the environment where the system performs its tasks is crucial for the correct operation of the Digital Twin like wind speed, direction, environmental temperature, humidity, among other variables relevant for each application. Once all the available information about the system is collected, the application of data Analytics like Principal Component analysis, signal denoising and detrending, average removals, descriptive and inferential statistics, and machine learning techniques are required to perform cleaning, filtering, and organization of the collected information, especially in big data scenarios with multiple agents interaction with unknown behaviors, where some data can be corrupted, missing, or is irrelevant to the system dynamics. 

\subsection{Step 3: Multidomain simulation}
In this step, the definition and configuration of the simulation models for representing the real system behavior are performed. In this case, the first task is defining the simulation domains related to the system. It means, defines the physical and constitutive laws that domain the system and select the computational tools to represent it. Usually, these domains include thermal, electrical, fluids, or digital components that can be simulated using multiple physics-based simulators like COMSOL, ANSYS, MSC-ADAMS, Matlab Simscape, among other multiphysics software packages. In some cases, the system model also incorporates discrete and algorithmic elements like task scheduling or event-based situations that can be managed using coding. Sometimes, when there is not enough information about some physical model of the system or that behavior cannot be characterized adequately, data-driven models of the system can be employed as a black box to represent that unknown behavior. Once the simulation domains are defined, and each subsystem interaction is built using the corresponding computational simulation tools, the next step consists of integrating each single simulation model to reproduce the system behavior. In some cases, all the subsystem models can be integrated using a single software package, but it often results in a co-simulation model, combining the capabilities of each individual software. It is important to highlight that if the Digital Twin application is designed to run in parallel with the real system, the computational cost can be high according to the multiphysics simulation packages employed; Therefore, a trade-off between the model level of details and computational performance should be considered. The initial runs of the multidomain simulation may run based on some ideal conditions of the represented system in order to verify the convergence and flow of the simulation environment. However, further matching is required to make the Digital Twin mimic the real system behavior.

\subsection{Step 4: DT assembly and Behavioral Matching}
\begin{figure}[h]
	\centering
	\includegraphics[width=0.45\textwidth, height=0.1\textheight]{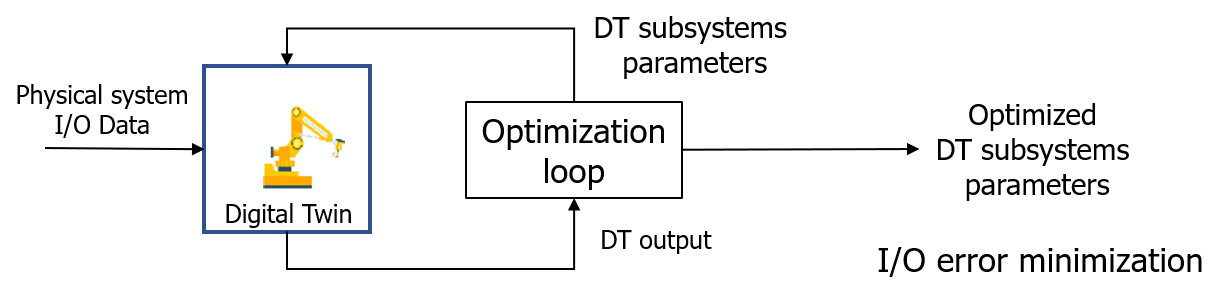}
	\caption{DT Behavioral Matching}
	\label{DT Behavioral Matching}
\end{figure}

In this step, a stable and operative multidomain simulation model of the system is available for the Digital Twin realization. Although, this model is operating under ideal conditions as stated before. So, a process called Behavioral Matching (BM) should be performed. It can be defined as a procedure to find the parameters of each subsystem composing the Digital Twin in order to fit its complete system dynamics, coinciding with the real state of the physical system. A description of the behavioral matching is presented in Fig.\ref{DT Behavioral Matching}. As can be observed, the complete Digital Twin model is feed with the input and output data from the real system, including the reference signals for control loops. Based on these data, the Digital Twin is set into an optimization loop, in which the main goal is to determine via optimal searching the unknown parameters that cannot be determined a-priori in each subsystem that compose the Digital Twin. This optimization loop keeps working until the coincidence between the input and output data streams of the system is reached with a certain tolerance rate or after a fixed number of iterations.
\par

There are some considerations to perform correct behavioral matching. Initially, the required level of detail of the Digital Twin has to be evaluated to known the complexity of the optimization process, considering that multidomain simulation models are complex and takes significant time to run a single simulation event. Also,  BM nature requires that not only the system output but also the input matches the real system operation. In addition, the BM is performed considering the complete multidomain simulation model, so that, choosing the most suitable optimization algorithm as well as set a cost function may be challenging. In that sense, metaheuristic methods like genetic algorithms and Cuckoo search can be an alternative, employing classic performance indices like ISE, ITAE, or IAE as cost function with weighted input and output signals as part of the function.

\subsection{Step 5: DT Validation and Deployment}
This is the final step on the Digital Twin implementation, performing the validation and simultaneous deployment with the real system. Initially, starting from the behavioral matching of the system, the Digital Twin response is calculated for different input/output data sets collected for the system. With this purpose, a supervisory interface has to be designed in order to perform the  Digital Twin offline and simultaneous execution of the system. Likewise, through the interface include the behavioral matching capabilities, combined with the analytics and fault detection modules, corresponding to the new enabling capabilities for the Digital Twin application. 

\section{Study case: real-time vision feedback infrared temperature uniformity control}
\subsection{Step 1: Target system definition}
\begin{figure}[h]
	\centering
	\includegraphics[width=0.3\textwidth, height=0.2\textheight]{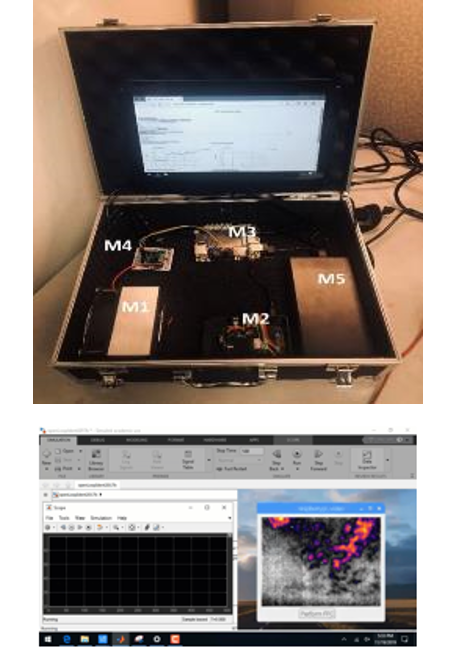}
	\caption{DT Study case: real-time vision feedback infrared temperature uniformity control}
	\label{DT Study case}
\end{figure}

The real-time vision feedback infrared temperature uniformity control presented in Fig.\ref{DT Study case} is employed in this paper as a study case for developing its Digital Twin based on the proposed methodology in Section III.
As shown in Fig.\ref{DT Study case}, The system is composed of a Peltier cell (M1) that works as a heating or cooling element, a thermal infrared camera (M2) acting as a temperature feedback sensor running on a Raspberry Pi and communicated using TCP/IP communication protocol, that allows performing temperature distribution measurement and control. An additional component of the system is the LattePanda board (M3), which runs Windows 10 64-bits and executes Matlab for local operation as well as the remote laboratory application for the NCS control system. To manage the power applied to the Peltier cell, an Arduino Leonardo board is used (M4), which via Pulse Width Modulation (PWM) controls the power driver. The platform is equipped with a battery (M5) that provides the power for all the components in the box with 4 hours of autonomy. This study case system can be fit into the second scenario corresponding to the system with a stable, controlled physical prototype. In this case, the uniform temperature control system is open and closed-loop stable, employing a PID controller with antwindup.
\subsection{Step 2: System documentation}
From step 1, the system is composed of four critical elements, the thermal infrared camera, the Peltier Thermoelectric module, the control unit (Lattepanda board), and the power driver, which operating specs are defined in \cite{Peltier,FLIR,Panda}. Table I and Table II present a brief summary of the critical properties for the power driver, the Peltier module and the thermal infrared camera, which are required for the steps of multidomain simulation and behavioral matching. In the case of the Peltier module, its properties differ among different manufacturers. Therefore, Behavioral Matching is required to determine the correct system parameters. More details about the system implementation and real test performed on the system can be found in \cite{Viola2019,Viola2019a}. 
\begin{table}[]
	\centering
	\caption{Brief thermal system documentation}
	\label{DTDocumentation}
	\begin{tabular}{@{}cc@{}}
		\toprule
		Component                                                                            & Features                                                                                                                                                 \\ \midrule
		\begin{tabular}[c]{@{}c@{}}FLIR lepton Thread\\ Infrared thermal Camera\end{tabular} & \begin{tabular}[c]{@{}c@{}}Wavelength: 8 to 14 $\mu$m\\ Resolution: 80x60 pixels\\ Accuracy: $\pm$ 0.5$^oC$\\\\\end{tabular} \\
		\begin{tabular}[c]{@{}c@{}}TEC1-12706 \\ Peltier Module\end{tabular}                 & \begin{tabular}[c]{@{}c@{}}$Q_{max}=50W$\\ $\Delta_{Tmax}=75^oC$\\ $I_{Max}=6.4A$\\ $V_{max}=16.4V$\\\\\end{tabular}         \\
		\begin{tabular}[c]{@{}c@{}}MC33926 DC\\ Power Driver\end{tabular}                    & \begin{tabular}[c]{@{}c@{}}Input: 0-5 V\\ Output: 0-12V\\ Peak Current: 5A\\\\\end{tabular}                                  \\
		Lattepanda board                                                                     & \begin{tabular}[c]{@{}c@{}}5 inch Windows 10 64 bits PC\\ Intel Atom $\mu p$\\ 4GB of RAM\\ Built-in Arduino Leonardo board\end{tabular}                 \\ \bottomrule
	\end{tabular}
\end{table}

\subsection{Step 3: multidomain Simulation}
\begin{figure}[h]
	\centering
	\includegraphics[width=0.45\textwidth, height=0.18\textheight]{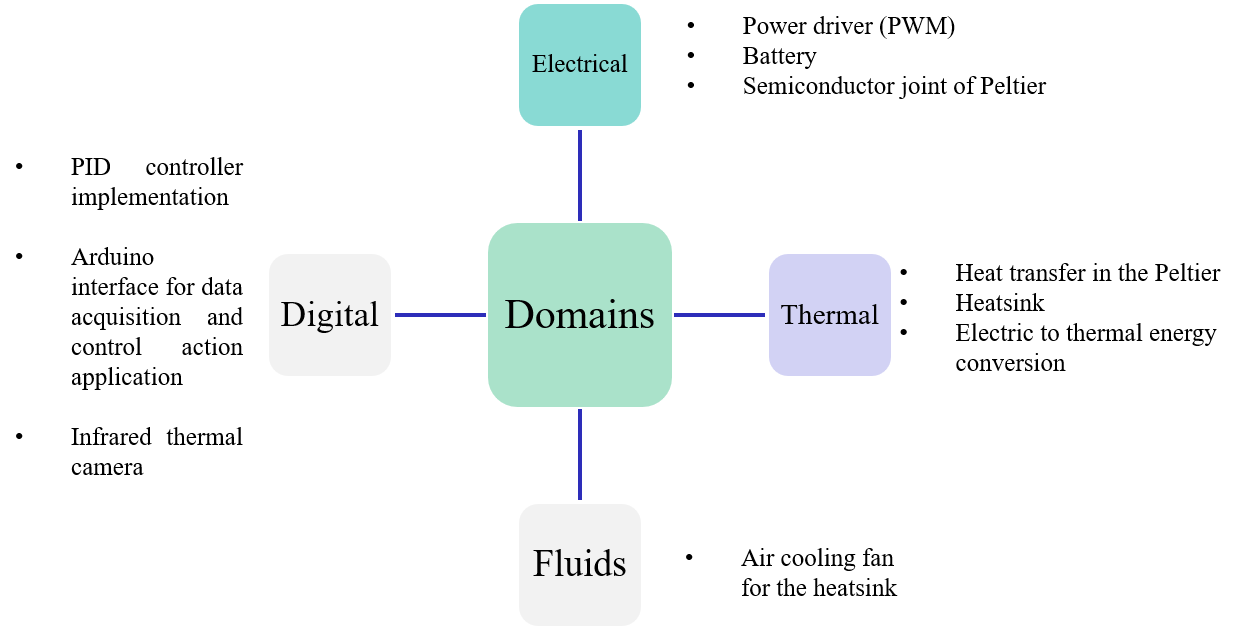}
	\caption{DT Study case: simulation domains}
	\label{DT simulation Domains}
\end{figure}

The study case is divided into four simulation domains presented in Fig.\ref{DT simulation Domains}. The first domain is the Electrical, composed by the power driver, the Battery, and the semiconductor joint on the Peltier module. The second one corresponds to the Thermal domain defined by the heat transfer produced between the Peltier hot and cold sides, the system surface and the surroundings, as well as the thermal properties of the heat sink. The third domain corresponds to the fluids part, given by the airflow pumped into the heat sink to keep its temperature constant. Finally, the fourth domain corresponds to the Digital Domain, composed by the PID control algorithm and the analog to digital interfaces to communicate the control side with the thermal system. Also, this simulation domain includes the behavior of the thermal infrared camera. 
\par
In this paper, the Electric, Thermal, and Digital domains will be replicated in the Digital Twin application. Matlab Simulink and Simscape electrical and thermal are employed as multidomain simulation packages to replicate the physical laws of the system as well as the PID control law employed. The complete multiphysics simulation model is presented in Fig.\ref{DTTotal}. A brief explanation of each domain is presented below.
\begin{figure}[h]
	\centering
	\includegraphics[width=0.49\textwidth, height=0.15\textheight]{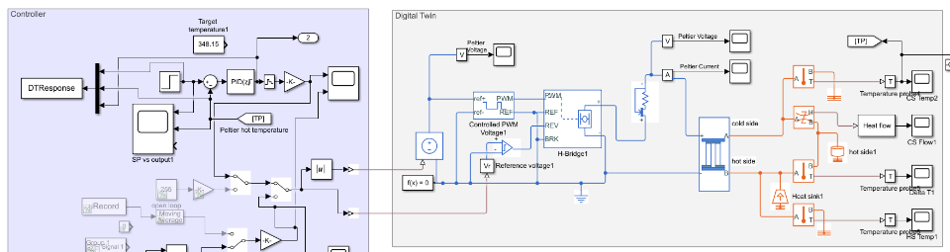}
	\caption{Assembled DT multidomain simulation}
	\label{DTTotal}
\end{figure}

\subsubsection{Electrical domain}
\begin{figure}[h]
	\centering
	\includegraphics[width=0.45\textwidth, height=0.15\textheight]{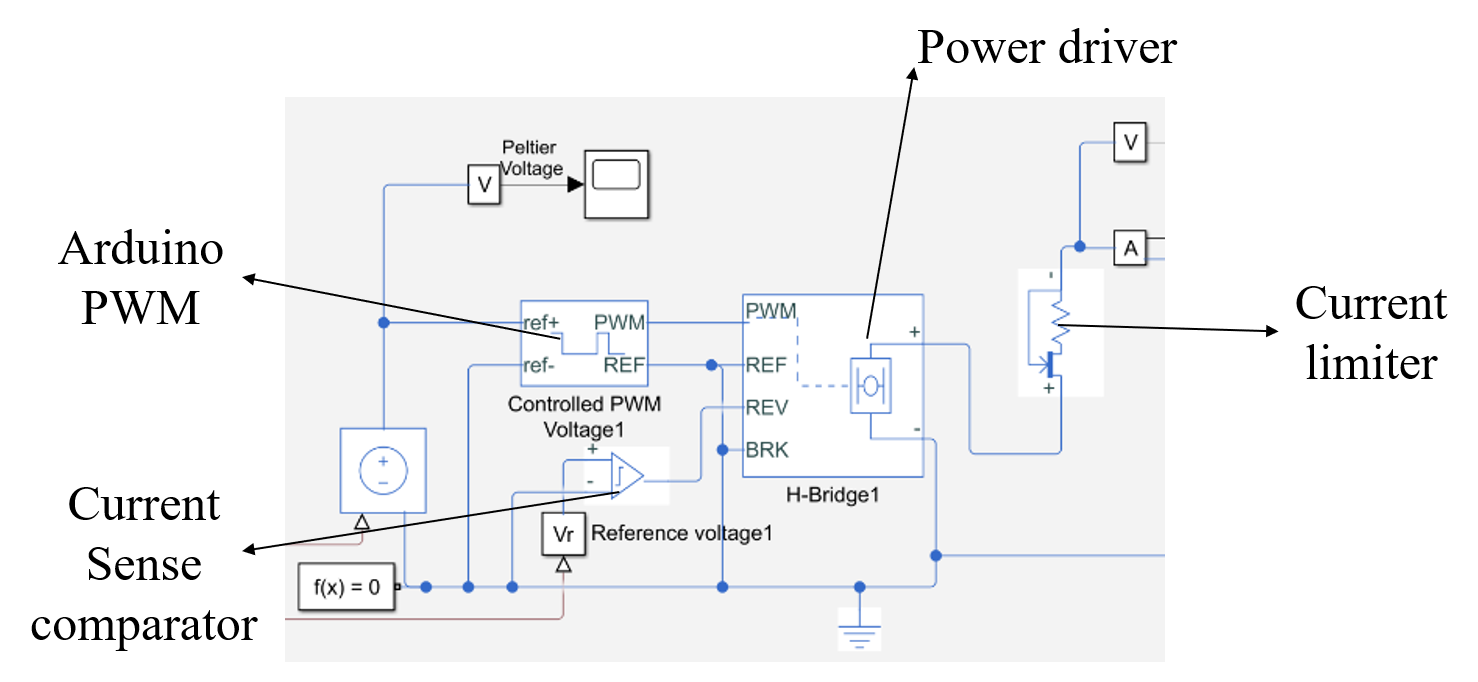}
	\caption{DT Study case: Electrical domain}
	\label{DTElectricalDomain}
\end{figure}

The electrical domain section of the DT model is shown in Fig.\ref{DTElectricalDomain}. This domain simulation is composed of the H-bridge to control the power flowing in the Peltier module via PWM using a controlled PWM voltage source. Also, a current sense comparator is included to set define the current flowing sense in the Peltier to alternate between heating and cooling behaviors. The PWM frequency is 500Hz given by the Arduino board used as the PWM control module in the real system, with the current sense threshold of 0.1v. The properties of the H-bridge are the same defined in Table\ref{DTDocumentation}.

\subsubsection{Thermal domain}
\begin{figure}[h]
	\centering
	\includegraphics[width=0.45\textwidth, height=0.14\textheight]{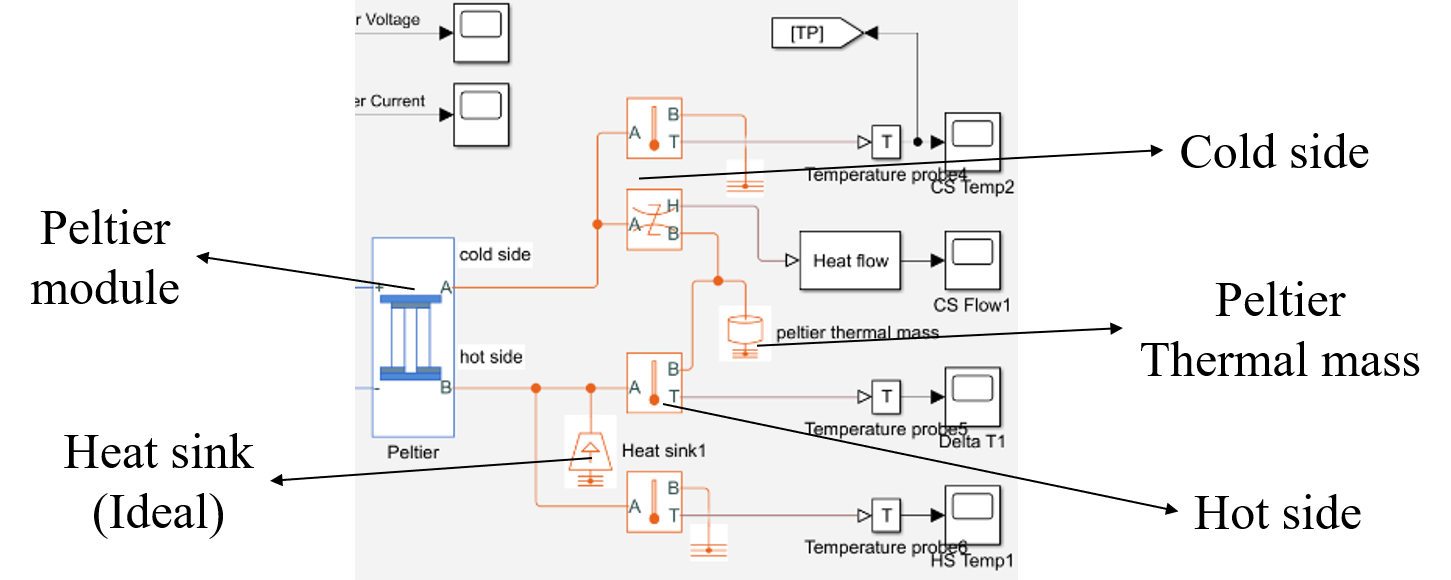}
	\caption{DT Study case: Thermal domain}
	\label{DTThermalDomain}
\end{figure}

\begin{figure}[h]
	\centering
	\includegraphics[width=0.4\textwidth, height=0.13\textheight]{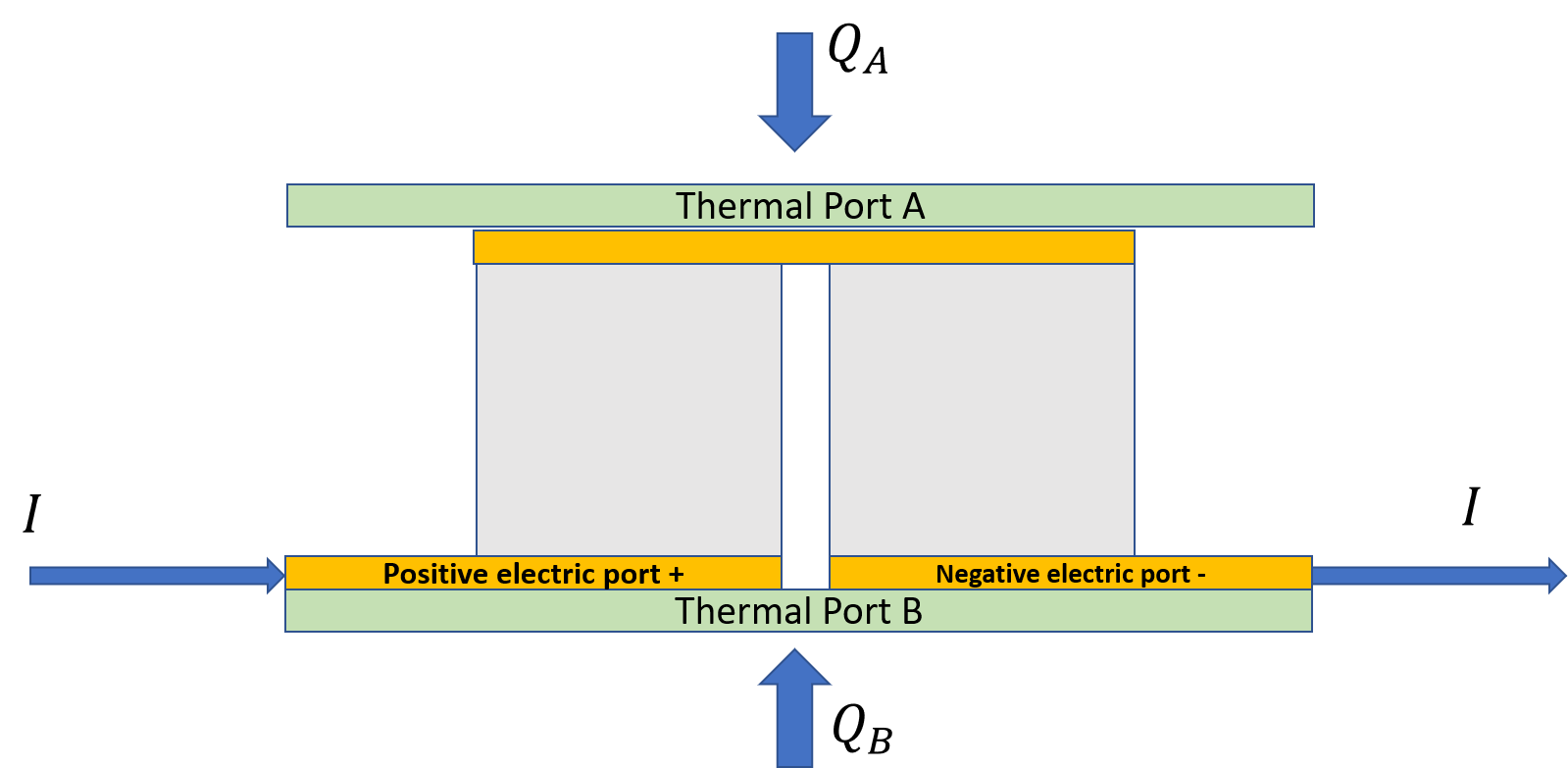}
	\caption{Peltier Thermoelectric module description}
	\label{PeltierGraph}
\end{figure}

: The thermal domain simulation is presented in Fig.\ref{DTThermalDomain}. As can be observed, the main element on this domain is the Peltier thermoelectric module, which is combined with a thermal mass and an ideal heat source to reproduce the real behavior of the Peltier. A description of the Simscape model of the Peltier thermoelectric module is presented in Fig.\ref{PeltierGraph}. It is composed of faces A (Hot side) and B (cold side) with an np semiconductor joint between A and B, through an applied electric current that will flow, producing a temperature difference. This  thermal interaction can be modeled using \eqref{peltierEqn1}-\eqref{peltierEqn3}, where $\alpha$: Seebeck coefficient, $R$ electrical resistance, $K$ thermal, conductance $T_A,T_B$: hot/cold side temperatures, $Q_A,Q_B$: hot/cold side thermal flow, and $I,V$ Peltier voltage and current. On the other hand, the thermal mass block represents the dynamic change of the heat flow $Q$ in the hot side of the Peltier, which behavior is given by \eqref{peltierEqn4}, where $C$ is the specific heat of the Peltier device and $m$ is the specific mass of the module. Finally, the heat sink is modeled by a constant temperature source at environment temperature, considering that its function keeps constant the temperature in the opposite face of the Peltier to produce the differential between A and B. Notice that $\alpha$,$R$,$K$, and $C$ are required to run the Digital Twin model, and initially can be derived from the Peltier datasheet. However, These values should be determined for the real system application. 

\begin{eqnarray}
Q_A=\alpha T_AI-\frac{1}{2}I^2R+K(T_A-T_B) \label{peltierEqn1}\\
Q_B=\alpha T_BI-\frac{1}{2}I^2R+K(T_B-T_A) \label{peltierEqn2} \\
V=\alpha(T_B-T_A)+IR  \label{peltierEqn3}\\
Q=Cm\frac{dT}{dt} \label{peltierEqn4}
\end{eqnarray}

\subsubsection{Digital domain}
\begin{figure}[h]
	\centering
	\includegraphics[width=0.45\textwidth, height=0.1\textheight]{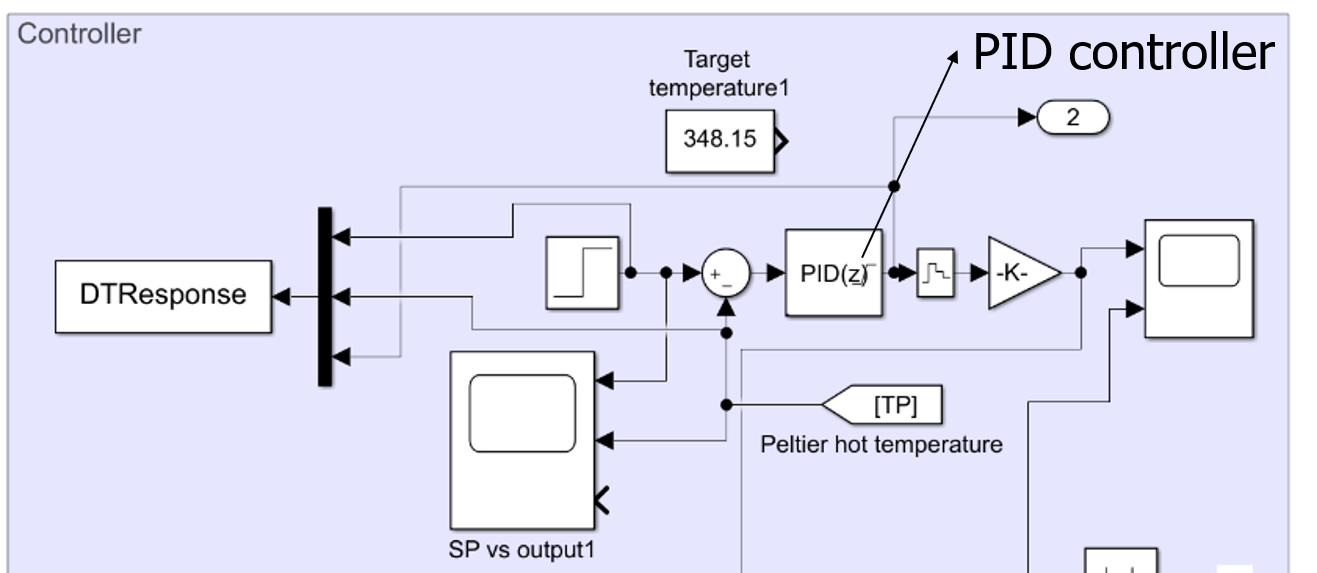}
	\caption{DT Study case: Digital domain}
	\label{DTDigitalDomain}
\end{figure}

The digital domain of the Digital Twin application is shown in Fig.\ref{DTDigitalDomain}. As can be observed, this domain includes the PID controller with anti-wind-up, the reference signal, and the control action to be applied as PWM. For this system, in particular, the physical implementation of this domain is performed using Hardware in the Loop simulation with Matlab Simulink. So, the behavior of these domains can be replicated with the best possible level of detail. The gains and configuration of the PID controller are described in detail at \cite{Viola2019,Viola2019a}.

%\subsubsection{Assembled DT}
%
%%\begin{figure}[h]
%%	\centering
%%	\includegraphics[width=0.45\textwidth, height=0.15\textheight]{images/DTTotal.png}
%%	\caption{Assembled DT multidomain simulation}
%%	\label{DTTotal}
%%\end{figure}

\subsection{Step 4: Behavioral matching}
\begin{figure}[h]
	\centering
	\includegraphics[width=0.5\textwidth, height=0.15\textheight]{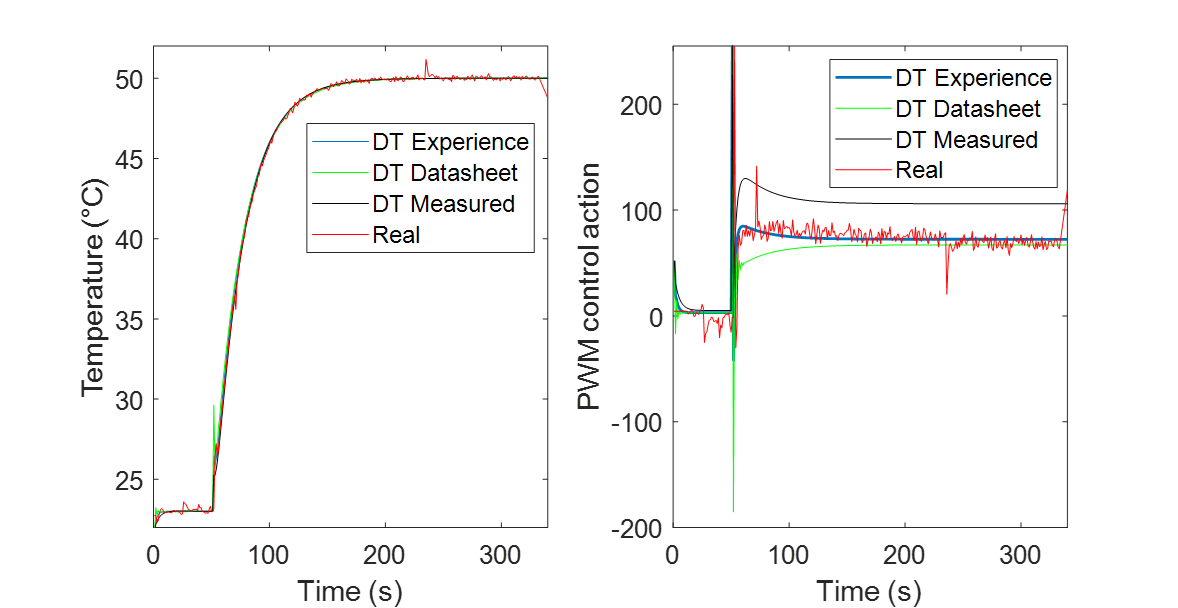}
	\caption{DT Response for different thermal parameters}
	\label{DTIOResponse}
\end{figure}

Based on the multidomain simulation model, the behavioral matching is performed to make the Digital Twin dynamics as similar as possible to the physical system. For the study case, the thermal domain parameters are the most relevant in the Digital Twin. So that, find the most accurate parameters Seebeck coefficient $\alpha$, electrical resistance $R$, thermal conductance $K$, and specific heat $C$  is required to obtain the best matching. There are different ways to determine these coefficients like direct measurement, as proposed in \cite{Kubov2016}, based on the datasheet information \cite{Peltier}, or by experience. Table\ref{DTBMPeltierParameters} shows the value for these parameters based on the methods described above.
\begin{table}[h]
	\centering
	\caption{Peltier Thermal parameters}
	\label{DTPeltierParameters}
	\begin{tabular}{@{}cccc@{}}
		\toprule
		Parameter & Datasheet    & Measurement & Experience    \\ \midrule
		$\alpha$  & 53 mv        & 40 mv       & 75 mv         \\
		$R$       & 1.8 $\Omega$ & 6 $\Omega$  & 2.90 $\Omega$ \\
		$K$       & 0.5555 K/W   & 0.3333 K/W  & 0.3808 K/W    \\
		C         & 15 J/K       & 15 J/K      & 31.4173 J/K   \\ \bottomrule
	\end{tabular}
\end{table}
The next step is to evaluate if these parameters make the Digital Twin behavior match with the real system input and output. For this purpose, a run of the physical system for a setpoint of $50^OC$ is performed, acquiring the control action, Peltier temperature output, and the reference setpoint applied. After that, the Digital Twin multidomain simulation is performed with the same reference signal and for each set of parameters in Table.\ref{DTPeltierParameters}. Figure.\ref{DTIOResponse} shows the Digital Twin responses for each set of parameters compared with the real system. As can be observed, each set of thermal parameters, the thermal output response looks similar but with different control input behaviors. It means that a systematic procedure to find the accurate parameters for the Peltier is required. One possible way is to disassemble the Peltier module and perform a characterization of each value through an experimental setup. However, in an industrial environment, this procedure may be impossible to perform. So that, in this paper, the behavioral matching is performed with a metaheuristic optimization approach employing a genetic algorithm (GA), which runs the Digital Twin model into an iterative loop to looks for the best values $\alpha$,$R$,$K$, and $C$, that makes the DT match with the real system response, it means, make the temperature output and the controller action match.

\begin{figure}[h]
	\centering
	\includegraphics[width=0.49\textwidth, height=0.12\textheight]{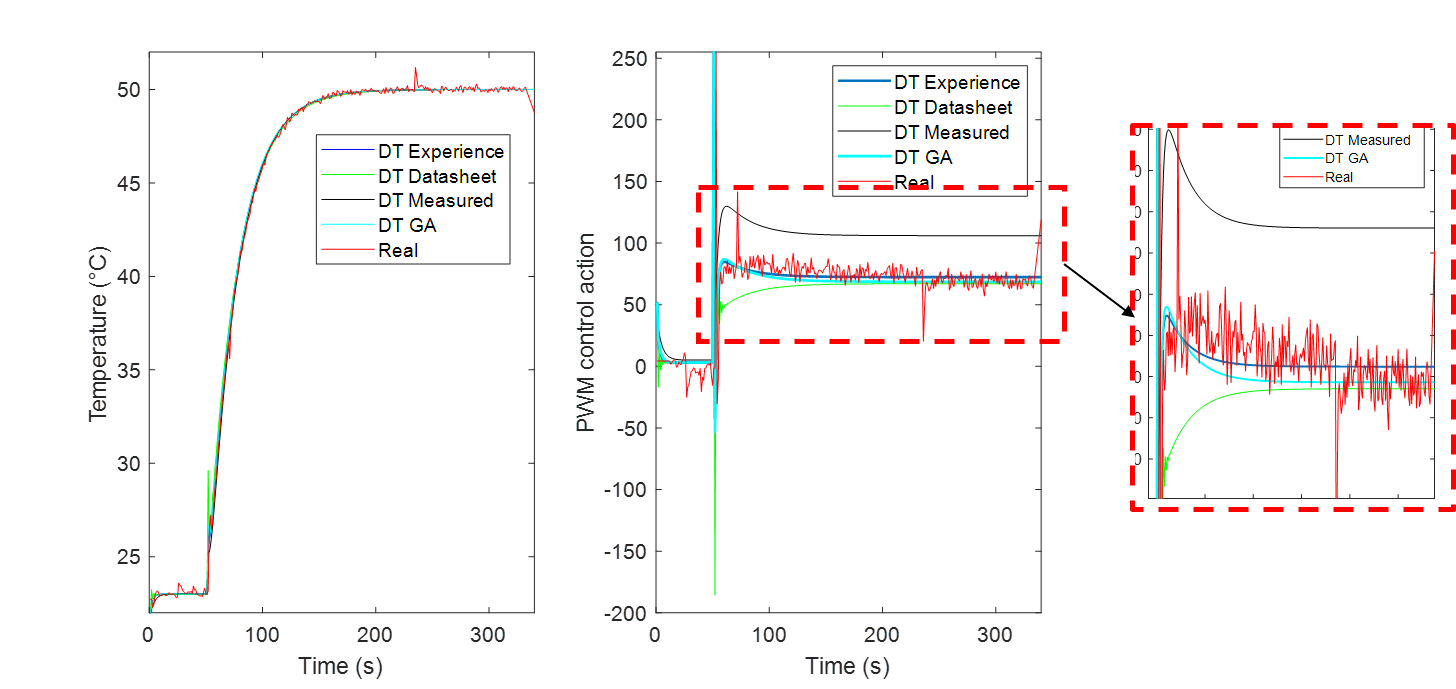}
	\caption{Behavioral matching results comparison}
	\label{DTBMvsEXP}
\end{figure}

The genetic algorithm employs the cost function given by \eqref{GACostFunction}, which is the sum of the quadratic error of the temperature output and the control action, where $J$ is the cost function value, $y_r$ the physical system output, $y_{DT}$ the digital twin output, $u_r$ the control action from the real system, and $u_{DT}$ the control action from Digital Twin, for $T$ seconds of duration. Also, the parameters constrains employed are $10mv\le \alpha\le200mv$, $1.8\le R\le6$, $0.2\le K\le0.833$, $15\le C\le30$. The GA is executed for 100 generations with an initial pool of three parents with a mutation probability of 90\%, and mutation rate of 2 features per crossing operation. The obtained parameters for the Peltier by the behavioral matching are $\alpha=35.8mv$, $R=3.35\Omega$ $K=0.2882 K/W$,  $C=15 J/K$. Figure.\ref{DTBMvsEXP} shows the response comparison among the Digital Twin for the Peltier parameters in Table.\ref{DTPeltierParameters} and those obtained with the behavioral matching. As can be observed, the optimization-based parameters improve the Digital Twin response, obtaining better fitness regarding the other sets due to these are closer to the real physical system parameters.

\begin{equation}
	min~ J=\int_{0}^{T} (y_r-y_{DT})^2+(u_r-u_DT)^2dt 
	\label{GACostFunction}
\end{equation}

%\begin{table}[h]
%	\centering
%	\caption{Obtained Peltier parameters from behavorial matching}
%	\label{DTBMPeltierParameters}
%	\begin{tabular}{@{}cc@{}}
%		\toprule
%		Parameter & GA            \\ \midrule
%		$\alpha$  & 35.8mv        \\
%		$R$       & 3.35 $\Omega$ \\
%		$K$       & 0.2882 K/W    \\
%		C         & 15 J/K        \\ \bottomrule
%	\end{tabular}
%\end{table}
%
%
%\begin{figure}[h]
%	\centering
%	\includegraphics[width=0.45\textwidth, height=0.15\textheight]{images/DTBMGA.png}
%	\caption{Behavorial matching results for DT system}
%	\label{DTBMGA}
%\end{figure}

\subsection{Step 5: DT validation and deployment}
\begin{figure}[h]
	\centering
	\includegraphics[width=0.45\textwidth, height=0.15\textheight]{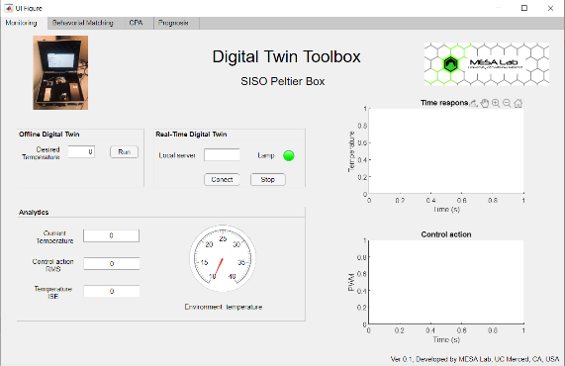}
	\caption{Digital Twin supervisory interface}
	\label{DTToolbox}
\end{figure}

Once the Digital Twin model passes through the behavioral matching process, it is ready for its deployment, running simultaneously with the physical system. For this reason, a supervisory interface, as well as a communication architecture, has to be defined to connect and monitor the real system with its Digital Twin. In this paper, the monitoring interface is implemented in Matlab using the app designer tool, which is shown in Fig.\ref{DTToolbox}. As can be observed, the interface offers the possibility of interacting with an offline version of the Digital Twin to verify its proper operation for different setpoints. Likewise, there is a panel for the real-time connection of the system with the Digital Twin.
\par
\begin{figure}[h]
	\centering
	\includegraphics[width=0.45\textwidth, height=0.1\textheight]{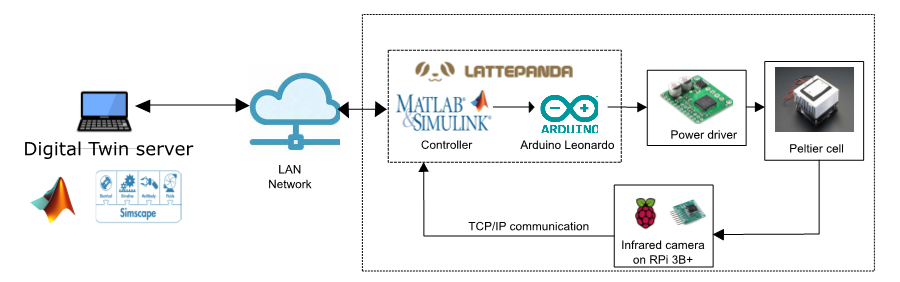}
	\caption{DT parallel deployment architecture}
	\label{DTDeployArchitecture}
\end{figure}
\begin{figure}[h]
	\centering
	\includegraphics[width=0.45\textwidth, height=0.12\textheight]{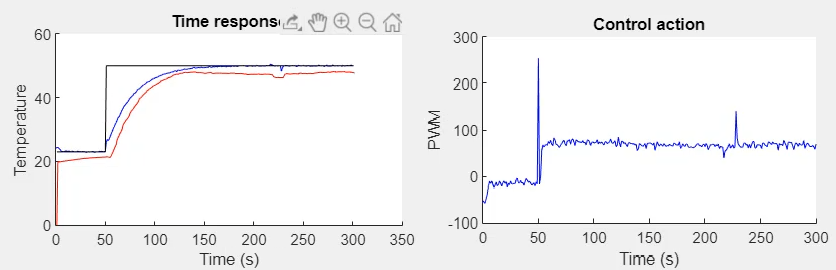}
	\caption{DT parallel deployment architecture}
	\label{DTOnlineStep}
\end{figure}

On the other hand, the communication interface among the DT and the system is presented in Fig.\ref{DTDeployArchitecture}. This interface connects the external computer where the Digital Twin is executed using a client/server configuration inside a local network with the real thermal system via TCP/IP protocol with a communication frequency of one second. Through this interface, the Digital Twin is feed in real-time with the same control action, output temperature, environmental temperature among other variables of the physical system to simulate its behavior and see how different is the Digital Twin running in real-time regarding the system. It is important to notice that, even with multiphysics simulation software like simscape, this Digital Twin implementation is able to handle the real-time simultaneous running. However, as the model increases its complexity, more computational power is required to run the Digital Twin in this way.
Figure \ref{DTOnlineStep} shows the results of the Real-time Digital Twin operation (blue line) with the real system (blue line), for a setpoint of $50^oC$. As can be observed, the Digital Twin replicates the real system behavior under real operation conditions. Also, it is important to notice that the environmental temperature has a significant influence over the Digital Twin performance, considering that it changes the starting point of the physical system. For this reason, this variable is measured and incorporated in the Digital Twin modeling and execution. A video recording of the Digital Twin interface operation can be visualized in https://youtu.be/acXTNmcCIYs.
\section{Conclusions and future works}
In this paper, a methodological framework has been proposed to build a Digital Twin for a closed-loop system real-time vision feedback infrared temperature uniformity control towards the implementation of Smart Control Engineering.
This framework employs five steps, which go from a detailed review of each component and subsystem, recreating the system behavior using multidomain simulation, adjusting to the real system using the behavioral matching technique to finish with the real-time interface between the Digital Twin and the physical system. Likewise, the obtained results show that the Digital Twin represents correctly the system behavior operating in parallel and offline modes. At this point, the Digital Twin of the real-time vision feedback infrared temperature uniformity control system is ready to incorporate all the enabling technologies required for the implementation of  Smart Control Engineering. As future works, the introduction of enabling technologies like fault detection and prognosis along with the Digital Twin is proposed. 
\bibliographystyle{ieeetr}
\bibliography{reference1}

\end{document}